\documentclass[11pt,a4paper]{article}
\usepackage[english]{babel}
\usepackage{amsmath,amsthm,amssymb,epsfig,latexsym}
\usepackage{color}
\usepackage{ulem}

\newcommand{\so}{\scriptscriptstyle \rm I}
\newcommand{\st}{\scriptscriptstyle \rm I\hspace{-1pt}I}

\newcommand{\la}{u}
\newcommand{\muu}{v}
\newcommand{\lac}{u^{\scriptscriptstyle C}}
\newcommand{\lab}{u^{\scriptscriptstyle B}}
\newcommand{\muc}{v^{\scriptscriptstyle C}}
\newcommand{\mub}{v^{\scriptscriptstyle B}}
\newcommand{\as}{\lambda}
\newcommand{\bla}{\bar u}
\newcommand{\bmu}{\bar v}
\newcommand{\blac}{\bar{u}^{\scriptscriptstyle C}}
\newcommand{\blab}{\bar{u}^{\scriptscriptstyle B}}
\newcommand{\bmuc}{\bar{v}^{\scriptscriptstyle C}}
\newcommand{\bmub}{\bar{v}^{\scriptscriptstyle B}}

\def\Izer{{\sf K}}
\newcommand{\be}[1]{\begin{equation}\label{#1}}
\newcommand{\ba}[1]{\begin{multline}\label{#1}}
\newcommand{\ee}{\end{equation}}
\newcommand{\ea}{\end{eqnarray}}

\newcommand{\num}{\\\rule{0pt}{20pt}}

\newcommand{\dis}{\displaystyle}

\newcommand{\diag}{\mathop{\rm diag}}
\newcommand{\tr}{\mathop{\rm tr}}

\def\Izer{{\sf K}}

\newtheorem{lemma}{Lemma}[section]

 \makeatletter
 \@addtoreset{equation}{section}
 \makeatother
 
\newcommand{\bea}{\begin{eqnarray}}
\newcommand{\eea}{\end{eqnarray}}
\newcommand{\cNr}{\raisebox{.21ex}{$\stackrel{\circ}{{\mathcal N}}$}}

\begin{document}
\begin{flushright}
LAPTH-053/12
\end{flushright}

\vspace{20pt}

\begin{center}
\begin{LARGE}
{\bf Form factors in $SU(3)$-invariant integrable models}
\end{LARGE}

\vspace{50pt}

\begin{large}
{S.~Belliard${}^a$, S.~Pakuliak${}^b$, E.~Ragoucy${}^c$, N.~A.~Slavnov${}^d$\footnote[1]{samuel.belliard@univ-montp2.fr, pakuliak@theor.jinr.ru, eric.ragoucy@lapp.in2p3.fr, nslavnov@mi.ras.ru}}
\end{large}

 \vspace{15mm}

${}^a$ {\it  Universit\'e Montpellier 2, Laboratoire Charles Coulomb,\\ UMR 5221,
F-34095 Montpellier, France}

\vspace{5mm}

${}^b$ {\it Laboratory of Theoretical Physics, JINR, 141980 Dubna, Moscow reg., Russia,\\
Moscow Institute of Physics and Technology, 141700, Dolgoprudny, Moscow reg., Russia,\\
Institute of Theoretical and Experimental Physics, 117259 Moscow, Russia}

\vspace{5mm}

${}^c$ {\it Laboratoire de Physique Th\'eorique LAPTH, CNRS and Universit\'e de Savoie,\\
BP 110, 74941 Annecy-le-Vieux Cedex, France}

\vspace{5mm}

${}^d$ {\it Steklov Mathematical Institute,
Moscow, Russia}

\end{center}

\vspace{4mm}



\begin{abstract}
We study $SU(3)$-invariant integrable models solvable by a nested algebraic Bethe ansatz.
We obtain determinant representations for form factors of diagonal entries of the
monodromy matrix. This representation can be used for the calculation of form factors
and correlation functions of the XXX $SU(3)$-invariant Heisenberg chain.
\end{abstract}

\vspace{1cm}

\section{Introduction}

Calculation of  form factors and correlation functions in the algebraic Bethe ansatz solvable models  \cite{FadST79,FadT79,BogIK93L,FadLH96,KulRes83}
is a very important  task. In many cases this can be reduced to the calculation of scalar products
of Bethe vectors. Recently, in the work \cite{BelPRS12b}, we obtained a determinant representation for a
particular case of the scalar product in $SU(3)$-invariant models. Using this representation, one can calculate
certain form factors in $SU(3)$-invariant Heisenberg chain.

In the present paper we extend this result and obtain determinant formulas for form factors of all
diagonal elements of the monodromy matrix. Let us be more precise. We consider models with $SU(3)$-invariant
$R$-matrix acting in the tensor product of two auxiliary spaces $V_1\otimes V_2$, where
$V_k\sim\mathbb{C}^3$, $k=1,2$:
 \be{R-mat}
 R(x,y)=\mathbf{I}+g(x,y)\mathbf{P},\qquad g(x,y)=\frac{c}{x-y}.
 \ee
In the above definition, $\mathbf{I}$ is the identity matrix in $V_1\otimes V_2$, $\mathbf{P}$ is the permutation matrix
that exchanges $V_1$ and $V_2$, and $c$ is a constant.

The monodromy matrix $T(w)$ satisfies the algebra
\be{RTT}
R_{12}(w_1,w_2)T_1(w_1)T_2(w_2)=T_2(w_2)T_1(w_1)R_{12}(w_1,w_2).
\ee
Equation \eqref{RTT} holds in the tensor product $V_1\otimes V_2\otimes\mathcal{H}$,
where $V_k\sim\mathbb{C}^3$, $k=1,2$, are the auxiliary linear spaces, and $\mathcal{H}$ is the Hilbert space of the Hamiltonian of the model under consideration. The $R$-matrix acts non-trivially in $V_1\otimes V_2$, the matrices $T_k(w)$ act non-trivially in
$V_k\otimes \mathcal{H}$.
The trace in the auxiliary space $V\sim\mathbb{C}^3$ of the monodromy matrix, $\tr T(w)$, is called the transfer matrix. It is a generating
functional of integrals of motion of the model. The eigenvectors of the transfer matrix are
called on-shell Bethe vectors (or simply on-shell vectors). They can be parameterized by a set of complex parameters
satisfying the Bethe equations (see section~\ref{S-N}).

Besides the standard monodromy matrix we also consider a twisted monodromy matrix
$\rho T(w)$ (see \cite{Kor82,IzeK84,KitMST05}), where
$\rho$ is a matrix such that its tensor square commutes with the $R$-matrix:
$[\rho_1\rho_2,R_{12}]=0$.  The operator $\tr \rho T(w)$ is called the twisted transfer matrix.
Its eigenvectors  are called twisted on-shell Bethe vectors (or simply twisted on-shell vectors).
Like the standard on-shell vectors, they can be parameterized by a set of complex parameters
satisfying the twisted Bethe equations (see section~\ref{S-N}). In our previous publication \cite{BelPRS12b} we considered a special case of the twist matrix $\rho=\diag(1,\kappa,1)$, where $\kappa$ was a complex number. Now we consider the general case of the diagonal twist matrix $\rho=\diag(\kappa_1,\kappa_2,\kappa_3)$. Below we will use the shorthand notation $\bar\kappa=\{\kappa_1,\kappa_2,\kappa_3\}$ and denote the twisted monodromy matrix by $T_{\bar\kappa}(w)$.

In this paper we obtain a determinant
representation for the scalar  product of a twisted on-shell vector and a standard on-shell vector. When the twist is general, this determinant
representation is not exact. It is valid only up to the terms of order $(\kappa_1/\kappa_3-1)^2$.  This precision,
however, allows us to obtain exact determinant formulas for form factors of $T_{ss}(w)$,
$s=1,2,3$. Using
these representations one can calculate form factors of diagonal operators in the $SU(3)$-invariant XXX Heisenberg
chain via the inverse scattering problem \cite{KitMaiT99,MaiTer00}. Indeed, if $E^{s,s}_m$, $s=1,2,3$, is an elementary
unit associated with the $m$-th site of the chain $\left(E^{s,s}\right)_{jk}=\delta_{js}\delta_{ks}$, then
 \be{gen-sol-T}
 E^{s,s}_m =(\tr T(0))^{m-1}  T_{ss}(0)(\tr T(0))^{-m}.
 \ee
Since the action of the transfer matrix on on-shell vectors is trivial, we see that the form factors of $E^{s,s}_m$ are proportional to the those of $T_{ss}$.

The article is organized as follows. In section~\ref{S-N} we introduce the model under consideration and describe
the notation used in the paper. In section~\ref{S-R} we give the main results. In section~\ref{S-SP-FF} we
explain how the twisted transfer matrix can be used for evaluation of form factors of the operators $T_{ss}(w)$.
Section~\ref{S-D} is devoted to the derivation of the results given in section~\ref{S-R}. Appendix~\ref{A-NF} contains the proof of an auxiliary lemma.

\section{Notation\label{S-N}}

We basically use the same notation and conventions as in the paper \cite{BelPRS12b}.

Besides the function $g(x,y)$ we also introduce  rational functions
\be{fg}
 f(x,y)=1+g(x,y)=\frac{x-y+c}{x-y}
\ee
and
\be{univ-not}
  h(x,y)=\frac{f(x,y)}{g(x,y)}=\frac{x-y+c}{c},\qquad  t(x,y)=\frac{g(x,y)}{h(x,y)}=\frac{c^2}{(x-y)(x-y+c)}.
\ee

Sets of variables are always denoted  by bars: $\bmu$, $\blac$ etc.
Individual elements of the sets are denoted by subscripts: $v_j$, $\lab_k$ etc. As a rule, the number of elements in the
sets is not shown explicitly in the equations; however we give these cardinalities in
special comments after the formulas.
We also use a special notation  for subsets with one element omitted  $\bla_j=\bla\setminus\la_j$, $\bmuc_k=\bmuc\setminus\muc_k$
and so on.

In order to avoid  formulas being too cumbersome we use shorthand notation for products of scalar
functions. Namely, if functions $g$, $f$, $h$, $t$, as well as $\lambda_2$  (see \eqref{Tjj}) depend
on sets of variables, this means that one should take the product over the corresponding set.
For example,
 \be{SH-prod}
 \lambda_2(\bar u)=\prod_{u_j\in\bar u} \lambda_2(u_j);\quad
 g(v_k, \bar w)= \prod_{w_j\in\bar w} g(v_k, w_j);\quad
 f(\bla_{\st},\bla_{\so})=\prod_{u_j\in\bla_{\st}}\prod_{u_k\in\bla_{\so}} f(u_j,u_k).
 \ee
In the last equation of \eqref{SH-prod} the set $\bar u$ is divided into two subsets
$\bla_{\so}$, $\bla_{\st}$, and the double product is taken with respect to all
$u_k$ belonging to $\bla_{\so}$ and all $u_j$ belonging to $\bla_{\st}$.

Now we pass to the description of Bethe vectors. We assume that the monodromy matrix
possesses a pseudovacuum vector $|0\rangle$ and a dual pseudovacuum vector $\langle0|$. These vectors
are annihilated by the operators $T_{jk}(w)$, where $j>k$ for  $|0\rangle$ and $j<k$ for $\langle0|$.
At the same time both vectors are eigenvectors for the diagonal entries of the monodromy matrix:
 \be{Tjj}
 T_{jj}(w)|0\rangle=\as_j(w)|0\rangle, \qquad   \langle0|T_{jj}(w)=\as_j(w)\langle0|,
 \ee
where $\as_j(w)$ are some scalar functions. In the framework of the generalized model, $\as_j(w)$ remain free functional parameters. Actually, it is always possible to normalize
the monodromy matrix $T(w)\to \as_2^{-1}(w)T(w)$ so as to deal only with the ratios
 \be{ratios}
 r_1(w)=\frac{\as_1(w)}{\as_2(w)}, \qquad  r_3(w)=\frac{\as_3(w)}{\as_2(w)}.
 \ee

Generic Bethe vectors are special polynomials in the operators $T_{jk}(w)$ with $j<k$ applied to $|0\rangle$.
Similarly, dual Bethe vectors are special polynomials in the operators $T_{jk}(w)$ with $j>k$ applied to $\langle0|$.
The procedure used to construct these polynomials was formulated in \cite{KulRes83} (see also \cite{TarVar93,BelRag08}).
Their explicit form  was found in \cite{BelPRS12c}. In  \cite{BelPRS12b} we denoted Bethe vectors and their dual
ones by $|\bla;\bmu\rangle$ and $\langle\bla;\bmu|$ respectively, stressing that they depend on two sets of variables
$\bla$ and $\bmu$. In this paper we use the notation $\mathbb{B}^{a,b}(\bla;\bmu)$ for Bethe vectors and
$\mathbb{C}^{a,b}(\bla;\bmu)$ for dual ones. These vectors differ from $|\bla;\bmu\rangle$ and $\langle\bla;\bmu|$
by the normalization
\be{Normalization}
\mathbb{B}^{a,b}(\blab;\bmub)=\frac{|\blab;\bmub\rangle}{f(\bmub,\blab)\as_2(\bmub)\as_2(\blab)},\qquad
\mathbb{C}^{a,b}(\blac;\bmuc)=\frac{\langle\blac;\bmuc|}{f(\bmuc,\blac)\as_2(\bmuc)\as_2(\blac)}.
\ee
We use here superscripts $B$ and $C$ in order to distinguish the sets of parameters
entering these two vectors. In other words, unless explicitly
specified, the variables $\{\blab, \bmub\}$ in $\mathbb{B}^{a,b}$ and
$\{\blac, \bmuc\}$ in $\mathbb{C}^{a,b}$ are not supposed to be equal.
The normalization \eqref{Normalization} is more convenient for the calculation of the action of the monodromy matrix
entries $T_{jk}(w)$ on Bethe vectors and dual ones \cite{BelPRS12c}. The superscripts $a$ and $b$ show the cardinalities
of the sets $\bla$ and $\bmu$: $\#\bla=a$, $\#\bmu=b$.

Bethe vectors and dual Bethe vectors are related by the anti-automorphism $^\dag$ defined by
$T_{ij}(w)^\dag=T_{ji}(w)$ and $|0\rangle^\dag=\langle0|$.

Below we will consider the scalar product of the on-shell vector and dual twisted on-shell  vector.
A generic Bethe vector  becomes an on-shell vector, if it is an eigenvector of the transfer matrix. Similarly the dual twisted on-shell  vector is an eigenvector of the twisted transfer matrix. We use the same notation, $\mathbb{B}^{a,b}(\blab;\bmub)$
and $\mathbb{C}^{a,b}(\blac;\bmuc)$, for on-shell vectors and dual ones, while we denote dual twisted on-shell  vectors by $\mathbb{C}_{\bar\kappa}^{a,b}(\blac;\bmuc)$ in order to stress that they are eigenvectors of $\tr T_{\bar\kappa}(w)$.
Then
\be{Left-act}
\begin{array}{l}
\tr T(w)\ \mathbb{B}^{a,b}(\blab;\bmub) = \tau(w|\blab,\bmub)\,\mathbb{B}^{a,b}(\blab;\bmub),\num
 \mathbb{C}_{\bar\kappa}^{a,b}(\blac;\bmuc)\ \tr T_{\bar\kappa}(w) = \tau_{\bar\kappa}(w|\blac,\bmuc)\,\mathbb{C}_{\bar\kappa}^{a,b}(\blac;\bmuc),
\end{array}
\ee
where
\be{tau-def}
\begin{array}{l}
\tau(w)\equiv\tau(w|\blab,\bmub)=r_1(w)f(\blab,w)+f(w,\blab)f(\bmub,w)+r_3(w)f(w,\bmub),\num
\tau_{\bar\kappa}(w)\equiv\tau_{\bar\kappa}(w|\blac,\bmuc)=\kappa_1r_1(w)f(\blac,w)+\kappa_2 f(w,\blac)f(\bmuc,w)
+\kappa_3r_3(w)f(w,\bmuc).
\end{array}
\ee
Hereby  the sets $\blab$ and $\bmub$  should satisfy the system of nested Bethe ansatz
equations \cite{KulRes83}
\be{AEigenS-1}
r_1(\lab_{j})=\frac{f(\lab_{j},\blab_{j})}{f(\blab_{j},\lab_{j})}f(\bmub,\lab_{j}),\qquad
r_3(\mub_{j})=\frac{f(\bmub_{j},\mub_{j})}{f(\mub_{j},\bmub_{j})}f(\mub_{j},\blab),
\ee
while the sets $\blac$ and $\bmuc$ satisfy the twisted system  of nested Bethe ansatz
equations
\be{ATEigenS-1}
r_1(\lac_{j})=\frac{\kappa_2}{\kappa_1}\frac{f(\lac_{j},\blac_{j})}{f(\blac_{j},\lac_{j})}f(\bmuc,\lac_{j}),
\qquad r_3(\muc_{j})=\frac{\kappa_2}{\kappa_3}\frac{f(\bmuc_{j},\muc_{j})}{f(\muc_{j},\bmuc_{j})}f(\muc_{j},\blac).
\ee
We recall that $\blac_{j}=\blac\setminus\lac_j$, $\blab_{j}=\blab\setminus\lab_j$ etc.

For further application it is useful to re-write the system of twisted equations in the logarithmic form. Define
\be{Phi-1}
\Phi_j=\log r_1(\lac_{j})-\log \left(\frac{f(\lac_{j},\blac_{j})}{f(\blac_{j},\lac_{j})}\right) -\log f(\bmuc,\lac_{j}),
\qquad j=1,\dots,a,
\ee
and
\be{Phi-2}
\Phi_{j+a}=\log r_3(\muc_{j})-\log \left(\frac{f(\bmuc_{j},\muc_{j})}{f(\muc_{j},\bmuc_{j})}\right)-\log f(\muc_{j},\blac),
\qquad j=1,\dots,b.
\ee
Then the system \eqref{ATEigenS-1} takes the form
\be{Log-TBE}
\begin{array}{l}
\Phi_j=\log\kappa_2-\log\kappa_1+2\pi i \ell_j,\qquad j=1,\dots,a,\\
\Phi_{j+a}=\log\kappa_2-\log\kappa_3+2\pi i m_j,\qquad j=1,\dots,b,
\end{array}
\ee
where $\ell_j$ and $m_j$ are some integers. The Jacobian of \eqref{Phi-1} and \eqref{Phi-2} is closely related to the
norm of the on-shell Bethe vector and the average values of the operators $T_{ss}(z)$ (see Section~\ref{S-R}).

To conclude this section we introduce the partition function of the six-vertex model with domain wall boundary conditions (DWPF) \cite{Kor82,Ize87}. This is one of the central object in the study of scalar products.  We denote the DWPF by
$\Izer_n(\bar x|\bar y)$. It depends on two sets of variables $\bar x$ and $\bar y$; the subscript shows that
$\#\bar x=\#\bar y=n$. The function $\Izer_n$ has the following determinant representation \cite{Ize87}
\begin{equation}\label{K-def}
\Izer_n(\bar x|\bar y)
=\Delta'_n(\bar x)\Delta_n(\bar y)h(\bar x,\bar y)
\det_n t(x_j,y_k),
\end{equation}
where $\Delta'_n(\bar x)$ and $\Delta_n(\bar y)$ are
\be{def-Del}
\Delta'_n(\bar x)
=\prod_{j>k}^n g(x_j,x_k),\qquad {\Delta}_n(\bar y)=\prod_{j<k}^n g(y_j,y_k).
\ee

\section{ Determinant expressions for the form factors\label{S-R}}

The form factors of the operators $T_{ss}(z)$ are defined as
 \be{SP-deFF}
 \mathcal{F}_{a,b}^{(s)}(z)\equiv\mathcal{F}_{a,b}^{(s)}(z|\blac,\bmuc;\blab,\bmub)=
 \mathbb{C}^{a,b}(\blac;\bmuc)T_{ss}(z)\mathbb{B}^{a,b}(\blab;\bmub),
 \ee
where both $\mathbb{C}^{a,b}(\blac;\bmuc)$ and $\mathbb{B}^{a,b}(\blab;\bmub)$ are on-shell
Bethe vectors\footnote[1]{%
For simplicity here and below we do not distinguish between vectors and dual vectors.}. One should distinguish two cases:
\begin{itemize}
\item
$\mathbb{C}^{a,b}(\blac;\bmuc)=
\bigl(\mathbb{B}^{a,b}(\blab;\bmub)\bigr)^\dagger$;
\item
 $\mathbb{C}^{a,b}(\blac;\bmuc)\ne
\bigl(\mathbb{B}^{a,b}(\blab;\bmub)\bigr)^\dagger$.
\end{itemize}

\subsection{The average value of $T_{ss}(z)$\label{ss-AV1}}

Here we consider the case $\mathbb{C}^{a,b}(\blac;\bmuc)=
\bigl(\mathbb{B}^{a,b}(\blab;\bmub)\bigr)^\dagger$, that is $\blab=\blac=\bla$ and  $\bmub=\bmuc=\bmu$.

First of all we define an  $(a+b)\times(a+b)$ matrix $\theta$ with the entries
\be{theta}
\theta_{j,k}=\left.\frac{\partial\Phi_j}{\partial \lac_k}\right|_{\blac=\bla\atop{\bmuc=\bmu}},\qquad k=1,\dots,a;\quad\text{and}\quad
\theta_{j,k+a}=\left.\frac{\partial\Phi_j}{\partial \muc_k}\right|_{\blac=\bla\atop{\bmuc=\bmu}},\qquad k=1,\dots,b,
\ee
where the $\Phi_j$ are given by  \eqref{Phi-1} and \eqref{Phi-2}.

Then we extend the matrix $\theta$ to an $(a+b+1)\times(a+b+1)$ matrix $\Theta^{(s)}$ with $s=1,2,3$, by adding one
row and one column
\be{Theta}
\begin{array}{l}
\Theta^{(s)}_{j,k}=\theta_{j,k}, \qquad j,k=1,\dots,a+b,\num
\Theta^{(s)}_{a+b+1,k}=\frac{\partial\tau(z|\bla,\bmu)}{\partial\la_k},\qquad k=1,\dots, a,\qquad
\Theta^{(s)}_{a+b+1,a+k}=\frac{\partial\tau(z|\bla,\bmu)}{\partial\muu_k},\qquad k=1,\dots, b,\num
\Theta^{(s)}_{j,a+b+1}=\delta_{s1}-\delta_{s2}\qquad j=1,\dots, a,\qquad
\Theta^{(s)}_{j+a,a+b+1}=\delta_{s3}-\delta_{s2}\qquad j=1,\dots, b,\num
\Theta^{(s)}_{a+b+1,a+b+1}=\left.\frac{\partial\tau_{\bar\kappa}(z|\blac,\bmuc)}{\partial\kappa_s}
\right|_{\blac=\bla\atop{\bmuc=\bmu}}.
\end{array}
\ee
Here the $\delta_{sk}$ are  Kronecker deltas. Notice that $\Theta^{(s)}$ depends on $s$ only in its last column.
Then the form factor $\mathcal{F}_{a,b}^{(s)}(z)$ is
\be{average-Tss}
\mathcal{F}_{a,b}^{(s)}(z|\bla,\bmu;\bla,\bmu)=H_{a,b}\det_{a+b+1}\Theta^{(s)},
\ee
where
\be{Hab}
H_{a,b}=(-1)^ac^{a+b}f(\bmu,\bla)\prod_{j,k=1\atop{j\ne k}}^af(\la_j,\la_k)
 \prod_{j,k=1\atop{j\ne k}}^b f(\muu_j,\muu_k).
 \ee

\subsection{Form factor of $T_{ss}(z)$ between different states\label{ss-FFDS1}}

If $\mathbb{C}^{a,b}(\blac;\bmuc)\ne
\bigl(\mathbb{B}^{a,b}(\blab;\bmub)\bigr)^\dagger$, then we introduce a row-vector $\Omega$ with the following components:
\be{def-Omega}
\begin{array}{l}
{\dis \Omega_k=\prod\limits_{\ell=1}^a(\lac_k-\lab_\ell)
\prod\limits_{\ell=1\atop{\ell\ne k}}^a(\lac_k-\lac_\ell)^{-1},\qquad k=1,\dots,a,}\num
{\dis \Omega_{a+k}=\prod\limits_{m=1}^b(\mub_k-\muc_m)
\prod\limits_{m=1\atop{m\ne k}}^b(\mub_k-\mub_m)^{-1},\qquad k=1,\dots,b.}
\end{array}
\ee
Obviously there exists an integer $p\in\{1,\dots,a+b\}$, such that $\Omega_p\ne 0$. Let $p$ be fixed.
Then for $j\ne p$ we define the entries $\mathcal{N}^{(s)}_{j,k}$ of the $(a+b)\times(a+b)$ matrix $\mathcal{N}^{(s)}$ as

 \be{FF-P11}
 \mathcal{N}^{(s)}_{j,k}= c\,g^{-1}(w_k,\blac)\,g^{-1}(\bmuc,w_k)
\frac{\partial \tau(w_k|\blac,\bmuc)}{\partial\lac_j},\qquad j=1,\dots,a,\quad j\ne p,
 \ee
and
 \be{FF-P22}
 \mathcal{N}^{(s)}_{a+j,k}=-c\,g^{-1}(\bmub,w_k )\,g^{-1}(w_k,\blab)
\frac{\partial \tau(w_k|\blab,\bmub)}{\partial\mub_j},\qquad j=1,\dots,b,\quad j\ne p.
 \ee
In these formulas one should set $w_k=\lab_k$ for $k=1,\dots,a$ and $w_{k+a}=\muc_k$ for $k=1,\dots,b$.

The $p$-th row has the following elements
\be{Np}
\mathcal{N}^{(s)}_{p,k}=h(\bmuc,w_k)h(w_k,\blab)Y^{(s)}_k,
\ee
where again $w_k=\lab_k$ for $k=1,\dots,a$ and $w_{k+a}=\muc_k$ for $k=1,\dots,b$, and
\be{Y1}
\begin{array}{l}
{\dis Y^{(s)}_k=c\,(\delta_{s1}-\delta_{s2})+(\delta_{s1}-\delta_{s3})\lab_k\left(1-\frac{f(\bmub,\lab_k)}{f(\bmuc,\lab_k)}\right),\qquad k=1,\dots,a;}\num
{\dis  Y^{(s)}_{a+k}=c\,(\delta_{s3}-\delta_{s2})+(\delta_{s1}-\delta_{s3})(\muc_k+c)\left(1-\frac{f(\muc_k,\blac)}{f(\muc_k,\blab)}\right),\qquad k=1,\dots,b.}
\end{array}
\ee
Then
\begin{multline}\label{FF-dif}
\mathcal{F}_{a,b}^{(s)}(z|\blac,\bmuc;\blab,\bmub)=\bigl(\tau(z|\blac;\bmuc)-\tau(z|\blab;\bmub)\bigr)\\
\times \Omega^{-1}_{p}t(\bmuc,\blab)
 \Delta'_a(\blac)\Delta_a(\blab)\Delta'_b(\bmuc)\Delta_b(\bmub)
 \det_{a+b}\mathcal{N}^{(s)}_{jk}.
 \end{multline}
Note that the number $p$ of the modified row of the matrix $\mathcal{N}^{(s)}$ is arbitrary. The only condition
is that $\Omega_p\ne 0$. It is worth mentioning that in the case considered in subsection~\ref{ss-AV1} the vector $\Omega$
becomes a zero vector. Therefore one cannot take the limit $\blac=\blab$ and $\bmuc=\bmub$ in  \eqref{FF-dif}. This should not
be surprising, because the sets $\{\blac,\bmuc\}$ and $\{\blab,\bmub\}$ are not generic complex numbers, but they are
some fixed solutions of Bethe equations. Evidently, one cannot consider the limit in which one solution goes
to another.

\section{Scalar product and form factors\label{S-SP-FF}}

Let $\tr T_{\bar\kappa}(z)$ be the twisted transfer matrix and $\tr T(z)$ be the standard transfer matrix.
Consider
\be{Qm}
Q_{\bar\kappa}(z)=\mathbb{C}_{\bar\kappa}^{a,b}(\blac;\bmuc) \bigl(\tr T_{\bar\kappa}(z)-\tr T(z)\bigr)\mathbb{B}^{a,b}(\blab;\bmub),
\ee
where $\mathbb{C}_{\bar\kappa}^{a,b}(\blac;\bmuc)$ and $\mathbb{B}^{a,b}(\blab;\bmub)$ are twisted and standard on-shell
vectors respectively.  Obviously
\be{Qm-0}
Q_{\bar\kappa}(z)=\mathbb{C}_{\bar\kappa}^{a,b}(\blac;\bmuc)\sum_{j=1}^3(\kappa_j-1)T_{jj}(z) \mathbb{B}^{a,b}(\blab;\bmub),
\ee
and therefore
\be{Qm-00}
\frac{d Q_{\bar\kappa}(z)}{d\kappa_s}\Bigl.\Bigr|_{\bar\kappa=1}=
\mathbb{C}_{\bar\kappa}^{a,b}(\blac;\bmuc)\Bigl.\Bigr|_{\bar\kappa=1}T_{ss}(z) \mathbb{B}^{a,b}(\blab;\bmub).
\ee
Here $\bar\kappa=1$ means that $\kappa_j=1$ for $j=1,2,3$.  Observe that after setting  $\bar\kappa=1$
the  vector $\mathbb{C}_{\bar\kappa}^{a,b}(\blac;\bmuc)$ turns into the standard on-shell vector $\mathbb{C}^{a,b}(\blac;\bmuc)$.
Hence, we obtain the form factor of $T_{ss}(z)$ in the r.h.s. of \eqref{Qm-00}
\be{Qm-FF}
\frac{d Q_{\bar\kappa}(z)}{d\kappa_s}\Bigl.\Bigr|_{\bar\kappa=1}=
\mathcal{F}^{(s)}(z|\blac,\bmuc;\blab,\bmub).
\ee

On the other hand
\be{Qm-1}
Q_{\bar\kappa}(z)=\bigl(\tau_{\bar\kappa}(z|\blac;\bmuc)-\tau(z|\blab;\bmub)\bigr)\;\mathbb{C}_{\bar\kappa}^{a,b}(\blac;\bmuc)\mathbb{B}^{a,b}(\blab;\bmub),
\ee
where $\tau_{\bar\kappa}(z|\blac;\bmuc)$ and $\tau(z|\blab;\bmub)$ are the eigenvalues of $\tr T_{\bar\kappa}(z)$  and $\tr T(z)$ respectively. Consider the case when $\mathbb{C}_{\bar\kappa}^{a,b}(\blac;\bmuc)\bigl.\bigr|_{\bar\kappa=1}=
\bigl(\mathbb{B}^{a,b}(\blab;\bmub)\bigr)^\dagger$, that is $\blac=\blab=\bla$ and $\bmuc=\bmub=\bmu$ at $\bar\kappa=1$. Then taking
the  derivative of \eqref{Qm-1} with respect to $\kappa_s$ at $\bar\kappa=1$ we find
\be{Qm-2}
\mathcal{F}^{(s)}(z|\bla,\bmu;\bla,\bmu)=
\|\mathbb{B}^{a,b}(\bla;\bmu)\|^2\;\frac{d \tau_{\bar\kappa}(z|\blac;\bmuc)}{d\kappa_s}\Bigl.\Bigr|_{\bar\kappa=1},
\ee
and  one should set $\blac=\bla$ and $\bmuc=\bmu$ after taking the derivative of $\tau_{\bar\kappa}(z|\blac;\bmuc)$
with respect to  $\kappa_s$.
Note that here we take total derivative with respect to  $\kappa_s$. Therefore, differentiating the eigenvalue
$\tau_{\bar\kappa}(z|\blac;\bmuc)$ we should also differentiate $\blac$ and $\bmuc$ with respect to  $\kappa_s$, as
these parameters implicitly depend on $\kappa_s$ through the twisted Bethe equations \eqref{ATEigenS-1}.

If $\mathbb{C}_{\bar\kappa}^{a,b}(\blac;\bmuc)\bigl.\bigr|_{\bar\kappa=1}\ne
\bigl(\mathbb{B}^{a,b}(\blab;\bmub)\bigr)^\dagger$, then the scalar product in the r.h.s. of \eqref{Qm-1} vanishes at
$\bar\kappa=1$ (as a scalar product of two different on-shell vectors), and we obtain
\be{Qm-3}
\frac{d Q_{\bar\kappa}(z)}{d\kappa_s}\Bigl.\Bigr|_{\bar\kappa=1}=\bigl(\tau(z|\blac;\bmuc)-\tau(z|\blab;\bmub)\bigr)\;
\frac{d}{d\kappa_s}\,
\left(\mathbb{C}_{\bar\kappa}^{a,b}(\blac;\bmuc)\mathbb{B}^{a,b}(\blab;\bmub)\right)\Bigl.\Bigr|_{\bar\kappa=1}.
\ee
Thus, the form factor of $T_{ss}(z)$ between two different on-shell vectors is proportional
to the $\kappa_s$-derivative of the scalar product of the twisted and standard on-shell
vectors
\be{Qm-4}
\mathcal{F}_{a,b}^{(s)}(z|\blac,\bmuc;\blab,\bmub)=\bigl(\tau(z|\blac;\bmuc)-\tau(z|\blab;\bmub)\bigr)\;
\frac{d}{d\kappa_s}\,
\left(\mathbb{C}_{\bar\kappa}^{a,b}(\blac;\bmuc)\mathbb{B}^{a,b}(\blab;\bmub)\right)\Bigl.\Bigr|_{\bar\kappa=1}.
\ee
Observe that after taking the $\kappa_s$-derivative of the scalar product one should set $\bar\kappa=1$.
Hence, for the calculation of form factors it is sufficient to compute the scalar product up to
the terms $(\kappa_i-1)(\kappa_j-1)$, where $i,j=1,2,3$.

\section{Calculation of the form factors \label{S-D}}

\subsection{Average value of $T_{ss}(z)$\label{ss-AV2}}
In this section we assume that $\blac=\blab=\bla$ and $\bmuc=\bmub=\bmu$ at $\bar\kappa=1$.

As we have shown in the previous section, the form factor $\mathcal{F}_{a,b}^{(s)}(z|\bla,\bmu;\bla,\bmu)$ is
equal to the norm of the on-shell vector  $\|\mathbb{B}^{a,b}(\bla;\bmu)\|^2$ multiplied by the derivative
of the twisted transfer matrix eigenvalue with respect to  $\kappa_s$ (see \eqref{Qm-2}).
The norm of the on-shell Bethe vector was calculated in \cite{Res86} (see also \cite{BelPRS12b})
\be{Norm}
\|\mathbb{B}^{a,b}(\bla;\bmu)\|^2=H_{a,b}\det_{a+b}\theta,
\ee
where $H_{a,b}$ is given by \eqref{Hab} and $\theta_{j,k}$ is defined in  \eqref{theta},
where one should set $\blac=\bla$ and $\bmuc=\bmu$.

The total derivative of $\tau_{\bar\kappa}(z|\blac,\bmuc)$ with respect to $\kappa_s$
at $\bar\kappa=1$ and $\blac=\bla$, $\bmuc=\bmu$ is
\be{tot-der}
\left.\frac{d\tau_{\bar\kappa}(z|\blac,\bmuc)}{d\kappa_s}\right|_{\bar\kappa=1}=
\left(\frac{\partial\tau_{\bar\kappa}(z|\blac,\bmuc)}{\partial\kappa_s}
+\sum_{\ell=1}^a\frac{\partial\tau(z|\bla,\bmu)}{\partial\la_\ell}\frac{d\lac_\ell}{d\kappa_s}
+\sum_{m=1}^b\frac{\partial\tau(z|\bla,\bmu)}{\partial\muu_m}\frac{d\muc_m}{d\kappa_s}\right)_{\bar\kappa=1}.
\ee
In order to compute derivatives $d\blac/d\kappa_s$ and $d\bmuc/d\kappa_s$
at $\bar\kappa=1$ we differentiate \eqref{Log-TBE}:
\be{system}
\begin{array}{l}
{\dis\sum_{\ell=1}^a \theta_{j,\ell}\frac{d \lac_\ell}{d\kappa_s}+
\sum_{m=1}^b \theta_{j,m}\frac{d \muc_m}{d\kappa_s}=\delta_{s2}-\delta_{s1},\qquad j=1,\dots,a}\num
{\dis\sum_{\ell=1}^a \theta_{j+a,\ell}\frac{d \lac_\ell}{d\kappa_s}+
\sum_{m=1}^b \theta_{j+a,m}\frac{d \muc_m}{d\kappa_s}=\delta_{s2}-\delta_{s3},\qquad j=1,\dots,b.}
\end{array}
\ee
From this system we find
\be{sol-system}
\begin{array}{l}
{\dis \frac{d \lac_j}{d\kappa_s}= (\delta_{s2}-\delta_{s1})\sum_{\ell=1}^a (\theta^{-1})_{j,\ell}+
(\delta_{s2}-\delta_{s3})\sum_{m=1}^b (\theta^{-1})_{j,m+a},}\num
{\dis \frac{d \muc_j}{d\kappa_s}= (\delta_{s2}-\delta_{s1})\sum_{\ell=1}^a (\theta^{-1})_{j+a,\ell}+
(\delta_{s2}-\delta_{s3})\sum_{m=1}^b (\theta^{-1})_{j+a,m+a}.}
\end{array}
\ee
Substituting  \eqref{sol-system} into \eqref{Qm-2} and \eqref{tot-der} we obtain
\begin{multline}\label{norm-tau}
\mathcal{F}_{a,b}^{(s)}(z|\bla,\bmu;\bla,\bmu)=H_{a,b} \det_{a+b}\theta \;\Biggl\{ \frac{\partial\tau_{\bar\kappa}(z|\blac,\bmuc)}{\partial\kappa_s}
\\
+\sum_{\ell=1}^a\frac{\partial\tau(z|\bla,\bmu)}{\partial\la_\ell}
\left[(\delta_{s2}-\delta_{s1})\sum_{\ell'=1}^a (\theta^{-1})_{\ell,\ell'}+
(\delta_{s2}-\delta_{s3})\sum_{m'=1}^b (\theta^{-1})_{\ell,m'+a}\right]\\
+\sum_{m=1}^b\frac{\partial\tau(z|\bla,\bmu)}{\partial\muu_m}
\left[(\delta_{s2}-\delta_{s1})\sum_{\ell'=1}^a (\theta^{-1})_{m+a,\ell'}+
(\delta_{s2}-\delta_{s3})\sum_{m'=1}^b (\theta^{-1})_{m+a,m'+a}\right]\Biggr\}_{\bar\kappa=1}.
 \end{multline}
Let $\widehat\theta_{j,k}$ be a cofactor of the matrix element $\theta_{j,k}$ in the matrix $\theta$.
Then $\widehat\theta_{j,k} =(\theta^{-1})_{k,j}\,\det_{a+b}\theta$, and \eqref{norm-tau} turns into
\begin{multline}\label{norm-tau1}
\mathcal{F}_{a,b}^{(s)}(z|\bla,\bmu;\bla,\bmu)=H_{a,b} \;\Biggl\{ \det_{a+b}\theta\;
\frac{\partial\tau_{\bar\kappa}(z|\blac,\bmuc)}{\partial\kappa_s}
\\
+\sum_{\ell=1}^a\frac{\partial\tau(z|\bla,\bmu)}{\partial\la_\ell}
\left[(\delta_{s2}-\delta_{s1})\sum_{\ell'=1}^a \widehat\theta_{\ell',\ell}+
(\delta_{s2}-\delta_{s3})\sum_{m'=1}^b  \widehat\theta_{m'+a,\ell}\right]\\
+\sum_{m=1}^b\frac{\partial\tau(z|\bla,\bmu)}{\partial\muu_m}
\left[(\delta_{s2}-\delta_{s1})\sum_{\ell'=1}^a  \widehat\theta_{\ell',m+a}+
(\delta_{s2}-\delta_{s3})\sum_{m'=1}^b  \widehat\theta_{m'+a,m+a}\right]\Biggr\}_{\bar\kappa=1}.
 \end{multline}

On the other hand, developing  $\det_{a+b+1}\Theta^{(s)}$ over the last row and the last column, one has
\begin{equation}\label{Dev-Theta}
\det_{a+b+1}\Theta^{(s)}=\Theta^{(s)}_{a+b+1,a+b+1}\det_{a+b}\theta-\sum_{j=1}^{a+b}\sum_{k=1}^{a+b}
\Theta^{(s)}_{j,a+b+1}\Theta^{(s)}_{a+b+1,k}\;\widehat\theta_{j,k}\;.
\end{equation}
Substituting here the explicit expressions \eqref{Theta} for the entries of the matrix $\Theta^{(s)}$, we immediately
reproduce \eqref{norm-tau1}. In this way we prove \eqref{average-Tss}.

Observe that
\be{sum-Th-s}
\sum_{s=1}^3\Theta^{(s)}_{j,a+b+1}=0,\quad\text{for}\quad j=1,\dots,a+b.
\ee
This implies
\begin{multline}\label{sum-FF-s}
\mathbb{C}^{a,b}(\bla;\bmu)\tr T(z)\mathbb{B}^{a,b}(\bla;\bmu)=H_{a,b} \det_{a+b}\theta
\sum_{s=1}^3\left.\frac{\partial\tau_{\bar\kappa}(z|\blac,\bmuc)}{\partial\kappa_s}\right|_{\blac=\bla\atop{\bmuc=\bmu}}\\
=\tau(z|\bla,\bmu)H_{a,b} \det_{a+b}\theta=\tau(z|\bla,\bmu)\|\mathbb{B}^{a,b}(\bla;\bmu)\|^2,
\end{multline}
as it should. One can also easily check that at $a=0$ or $b=0$, equation \eqref{average-Tss}
reproduces known results for $SU(2)$ form factors  \cite{KitMaiT99}.

\subsection{The scalar product of twisted and standard on-shell vectors\label{ss-SP-TS}}

In order to compute the form factor of $T_{ss}(z)$ between different on-shell Bethe vectors we should calculate
the scalar product of the twisted on-shell vector and the standard on-shell vector (see \eqref{Qm-4}).
The main steps of this derivation almost literally repeat the ones described in the work \cite{BelPRS12b} for
the particular case of the twist matrix. We start with the general formula for the scalar product of two
Bethe vectors obtained by N. Reshetikhin in the work \cite{Res86}.  Then we successively take the sums
over partitions of the arguments of Bethe vectors. The reader can find the details of this very onerous derivation
in \cite{BelPRS12b}. The only essential difference is that now we need a generalization of lemma~6.3 of
\cite{BelPRS12b}.

\begin{lemma}\label{New-Lemma}
Let $\zeta$ be a constant. Define $G_n(\zeta)$ as
\be{def-Gg}
G_n(\zeta)=\sum \zeta^{n_{\st}}f(\bar\xi_{\so},\bar\xi_{\st})f(\bar\eta_{\st},\bar\eta_{\so})\Izer_{n_{\so}}(\bar\eta_{\so}|\bar\xi_{\so})
\Izer_{n_{\st}}(\bar\xi_{\st}+c|\bar\eta_{\st}),
\ee
where $n=n_{\so}+n_{\st}$, and the sum is taken over all partitions of the set $\bar\eta$ into subsets $\bar\eta_{\so},\bar\eta_{\st}$ and the set $\bar\xi$ into subsets $\bar\xi_{\so},\bar\xi_{\st}$ with cardinalities $\#\bar\eta_{\so}=\#\bar\xi_{\so}=n_{\so}$,  $0\leq n_{\so}\leq n$,
and $\#\bar\eta_{\st}=\#\bar\xi_{\st}=n_{\st} = n-n_{\so}$. The functions $\Izer_{n_{\so}}$ and $\Izer_{n_{\st}}$ are the DWPF \eqref{K-def}. Then
\be{res-Gg}
G_n(\zeta)=(-1)^n\zeta^{\frac{\bar\eta-\bar\xi}c}\; t(\bar\xi,\bar\eta)h(\bar\eta,\bar\eta)h(\bar\xi,\bar\xi)+
O\bigl((\zeta-1)^2\bigr),
\ee
where we have used the shorthand notation
\be{sh-not}
\zeta^{\frac{\bar\eta-\bar\xi}c}=\prod_{j=1}^n\zeta^{\frac{\eta_j-\xi_j}c}.
\ee
\end{lemma}

The proof is given in appendix~\ref{A-NF}.

It turns out that in our case, $\zeta=\kappa_1/\kappa_3$.
In the work \cite{BelPRS12b} we considered the case $\kappa_1=\kappa_3$. Therefore we succeeded in calculating the
sum \eqref{def-Gg} exactly. In the case of the general twist matrix we have $\kappa_1\ne \kappa_3$, and hence,
$\zeta\ne1$. Thus, generically the function $G_n(\zeta)$ is a polynomial in $\zeta$. Using lemma~(6.1) of
\cite{BelPRS12b} one can take the
sum in \eqref{def-Gg} with respect to the partitions of one set of variables, for instance,
\be{ParSum-Gg}
G_n(\zeta)=\sum \zeta^{n_{\st}}(-1)^{n_{\so}}f(\bar\xi_{\so},\bar\xi_{\st})f(\bar\eta,\bar\xi_{\so})
\Izer_{n}(\{\bar\xi_{\so}-c,\bar\xi_{\st}+c\}|\bar\eta).
\ee
Here the sum is taken only over  partitions of  the set $\bar\xi$ into subsets $\bar\xi_{\so},\bar\xi_{\st}$. However
it is doubtful that further simplifications of  equation \eqref{ParSum-Gg} are possible. This is a serious obstacle
for the derivation of a determinant representation for the scalar product involving twisted on-shell vectors
with a general twist.

On the other hand, in order to calculate form factors, we should find only the first $\kappa_s$-derivatives of the scalar product at $\bar\kappa=1$. Therefore we do not need an exact result for $G_n(\zeta)$, since the terms $O\bigl((\zeta-1)^2\bigr)$
are not relevant.

As we have pointed out, in all  other respects the derivation of the determinant representation for the scalar
product of twisted and standard on-shell Bethe vectors literally repeats the derivation described in
\cite{BelPRS12b}. The result reads
 \begin{equation}\label{fin0}
 \mathbb{C}_{\bar\kappa}^{a,b}(\blac;\bmuc)\mathbb{B}^{a,b}(\blab;\bmub)= t(\bmuc,\blab)
 \Delta'_a(\blac) \Delta'_b(\bmub)
 {\Delta}_a(\blab){\Delta}_b(\bmuc)\det_{a+b}\mathcal{N}+O\bigl((\kappa_3/\kappa_1-1)^2\bigr),
 \end{equation}
where $\Delta'$ and $\Delta$ are defined in \eqref{def-Del}. In order to describe the
$(a+b)\times(a+b)$ matrix $\mathcal{N}$ we first introduce a column-vector $\widehat{\mathcal{N}}_j(w)$
with the components
 \be{SP-P11}
\widehat{\mathcal{N}}_j(w)= c\,g^{-1}(w,\blac)g^{-1}(\bmuc,w)
\frac{\partial \tau_{\bar\kappa}(w|\blac,\bmuc)}{\partial\lac_j},\qquad j=1,\dots,a,
 \ee
and
 \be{SP-P22}
\widehat{\mathcal{N}}_{j+a}(w)=-c\,g^{-1}(\bmub,w )g^{-1}(w,\blab)
\frac{\partial \tau(w|\blab,\bmub)}{\partial\mub_j},\qquad j=1,\dots,b.
 \ee
Then
 \begin{equation}\label{block-matrix}
\begin{array}{l}
{\dis \mathcal{N}_{j,k}=\widehat{\mathcal{N}}_{j}(\lab_k),\qquad j,k=1,\dots,a;}\num
{\dis \mathcal{N}_{j,k}=\widehat{\mathcal{N}}_{j}(\muc_k)\left(\frac{\kappa_3}{\kappa_1}\right)^{\muc_{k}/c},
\qquad j=1,\dots,a,\quad k=a+1,\dots,b;}\num
{\dis \mathcal{N}_{j,k}=\widehat{\mathcal{N}}_{j}(\lab_k)\left(\frac{\kappa_3}{\kappa_1}\right)^{-\lab_{k}/c},
\qquad j=a+1,\dots,b,\quad k=1,\dots,a;}\num
{\dis \mathcal{N}_{j,k}=\widehat{\mathcal{N}}_{j}(\muc_k),\qquad j,k=a+1,\dots,b.}
\end{array}
\ee
Comparing the entries of the matrix \eqref{block-matrix} with the ones obtained in \cite{BelPRS12b} one can see
additional factors $\left(\kappa_3/\kappa_1\right)^{\muc_{k}/c}$ and
$\left(\kappa_3/\kappa_1\right)^{-\lab_{k}/c}$ in the off-diagonal blocks. These terms are
due to the factor $\zeta^{(\bar\eta-\bar\xi)/c}$ in lemma~\ref{New-Lemma}.

\subsection{The form factor of $T_{ss}(z)$ between different states\label{ss-FFDS2}}

In order to obtain form factors one has to take $\kappa_s$-derivatives of the scalar product at $\bar\kappa=1$.
Taking into account that the parameters $\blac$ and $\bmuc$  depend on $\bar\kappa$ through the twisted Bethe
equations, it might be rather difficult to obtain an explicit expression for the derivatives of $\det_{a+b}\mathcal{N}$.
However, as  was shown in  \cite{BelPRS12b}, the matrix  $\mathcal{N}$ has an eigenvector with zero eigenvalue
at $\bar\kappa=1$. This fact can be used for significant simplification of our calculations.

The components of the zero eigenvector are given by \eqref{def-Omega}. If $\mathbb{C}^{a,b}(\blac;\bmuc)\ne
\bigl(\mathbb{B}^{a,b}(\blab;\bmub)\bigr)^\dagger$ then this vector has at least
one non-zero component, say $\Omega_p$. Then we can add to the $p$-th row of the matrix $\mathcal{N}$
all other rows multiplied by the coefficients $\Omega_j/\Omega_{p}$. Such sums were already calculated in
\cite{BelPRS12b}.
Then  the $p$-th row is modified as follows:
\be{LR-l}
\mathcal{N}_{p,k}=c\Omega^{-1}_{p}h(\bmuc,\lab_k)h(\lab_k,\blab)\left[\frac{f(\bmub,\lab_k)}{f(\bmuc,\lab_k)}
\left(1-\left(\frac{\kappa_1}{\kappa_3}\right)^{\lab_k/c}\right)
+\left(\frac{\kappa_1}{\kappa_3}\right)^{\lab_k/c}-\frac{\kappa_2}{\kappa_1}\right],
\ee
for $k=1,\dots,a$, and
\be{LR-r}
\mathcal{N}_{p,a+k}=c\Omega^{-1}_{p}h(\bmuc,\muc_k)h(\muc_k,\blab)\left[\frac{f(\muc_k,\blac)}{f(\muc_k,\blab)}
\left(\frac{\kappa_2}{\kappa_1}\left(\frac{\kappa_3}{\kappa_1}\right)^{\muc_k/c}-\frac{\kappa_2}{\kappa_3}\right)
+1-\frac{\kappa_2}{\kappa_1}\left(\frac{\kappa_3}{\kappa_1}\right)^{\muc_k/c}\right],
\ee
for $k=1,\dots,b$.

Obviously $\mathcal{N}_{p,k}=0$ at $\bar\kappa=1$.
Hence, when we take derivatives with respect to $\kappa_s$ we have to differentiate only the $p$-th row, setting
$\bar\kappa=1$ everywhere else. We also do not need to take derivatives of $\blac$ and $\bmuc$
with respect to $\kappa_s$, since they produce zero contributions at $\bar\kappa=1$. Thus, differentiating \eqref{LR-l},
\eqref{LR-r} with respect to $\kappa_s$ we arrive at \eqref{Y1}. In all other rows of the matrix $\mathcal{N}$
we simply set $\bar\kappa=1$. In this way we reproduce equations \eqref{FF-P11} and \eqref{FF-P22}.

Observe that
\be{sum-Y-s}
\sum_{s=1}^3 Y^{(s)}_{k}=0,\quad\text{for}\quad k=1,\dots,a+b,
\ee
where $Y^{(s)}_{k}$ are given by  \eqref{Y1}. Hence,
\begin{equation}\label{sum-FF1-s}
\mathbb{C}^{a,b}(\blac;\bmuc)\tr T(z)\mathbb{B}^{a,b}(\blab;\bmub)=0,
\end{equation}
as it should, if $\mathbb{C}^{a,b}(\blac;\bmuc)\ne
\bigl(\mathbb{B}^{a,b}(\blab;\bmub)\bigr)^\dagger$. Known formulas for form factors in $SU(2)$ case  \cite{KitMaiT99} also can be obtained from \eqref{FF-dif} by setting $a=0$ or $b=0$.

\section*{Conclusion}

We have mentioned already that the problem of calculation of form factors and correlation functions
in the framework of the algebraic Bethe ansatz can be reduced to that of the calculation of scalar products
of Bethe vectors. In particular, determinant representations for scalar products of on-shell
vectors and generic Bethe vectors play a very important role. Such determinant representations for
$\frak{gl}_2$-based models are known \cite{Sla89,Sla07} and they were used for analytical \cite{KitMT00,KitMST02,KitKMST09b,GohKS04,GohKS05,SeeBGK07} and numerical \cite{CauHM05,PerSCHMWA06,PerSCHMWA07,CauCS07} analysis
of correlation functions. However in the case of higher rank algebras the situation is more
involved. We argued in \cite{BelPRS12b} that in the $SU(3)$ case a determinant representation for the scalar product of an on-shell
vector and arbitrary Bethe vector is hardly to be found. In the present paper we obtained an additional argument in favor of
this conjecture. Indeed, even in the particular case of a Bethe vector, namely a twisted on-shell vector, we were able
to obtain the determinant representation for the scalar product only up to the terms of order $(\kappa_1/\kappa_3-1)^2$.
In order to obtain an exact result one should obtain a closed expression for the function $G_n(\zeta)$ defined
in lemma~\ref{New-Lemma}.

Nevertheless we succeeded in finding single determinant representations for form factors of $T_{ss}(w)$. These matrix elements describe
form factors of diagonal operators in the $SU(3)$-invariant Heisenberg chain. For complete a description one should
obtain determinant formulas for the matrix elements of $T_{jk}(w)$ with $j\ne k$. We hope to study this problem in our further
publication.

\section*{Acknowledgements}
Work of S.P. was supported in part by RFBR grant 11-01-00980-a, grant
of Scientific Foundation of NRU HSE 12-09-0064 and grant of
FASI RF 14.740.11.0347. E.R. was supported by ANR Project
DIADEMS (Programme Blanc ANR SIMI1 2010-BLAN-0120-02).
N.A.S. was  supported by the Program of RAS Basic Problems of the Nonlinear Dynamics,
RFBR-11-01-00440, RFBR-11-01-12037-ofi-m, SS-4612.2012.1.

\appendix

\section{Proof of lemma~\ref{New-Lemma}\label{A-NF}}

\begin{lemma}\label{New-Lemma1}
Let $\gamma$ be a constant. Define $\tilde G_n(\gamma)$ as
\be{def-Gg1}
\tilde G_n(\gamma)=\gamma G_n(1)+\left.\frac{d G_n}{d\zeta}\right|_{\zeta=1}=\sum (\gamma+n_{\st})f(\bar\xi_{\so},\bar\xi_{\st})f(\bar\eta_{\st},\bar\eta_{\so})\Izer_{n_{\so}}(\bar\eta_{\so}|\bar\xi_{\so})
\Izer_{n_{\st}}(\bar\xi_{\st}+c|\bar\eta_{\st}),
\ee
where $n=n_{\so}+n_{\st}$, and the sum is taken over all partitions of the set $\bar\eta$ into subsets $\bar\eta_{\so},\bar\eta_{\st}$ and the set $\bar\xi$ into subsets $\bar\xi_{\so},\bar\xi_{\st}$ with cardinalities $\#\bar\eta_{\so}=\#\bar\xi_{\so}=n_{\so}$,
$0\leq n_{\so}\leq n$, and $\#\bar\eta_{\st}=\#\bar\xi_{\st}=n_{\st} = n-n_{\so}$.
The functions $\Izer_{n_{\so}}$ and $\Izer_{n_{\st}}$ are the DWPF \eqref{K-def}.

Then
\be{res-Gg1}
\tilde G_n(\gamma)=(-1)^n t(\bar\xi,\bar\eta)h(\bar\eta,\bar\eta)h(\bar\xi,\bar\xi)\left(\gamma+\sum_{i=1}^ng^{-1}(\eta_i,\xi_i)\right).
\ee
\end{lemma}

{\sl Proof.} The proof is similar to that of lemma~6.3 of
\cite{BelPRS12b}. The function $\tilde G_n(\gamma)$ is a rational function of
$\bar\eta$ and $\bar\xi$. It is symmetric in $\bar\eta$ and symmetric in $\bar\xi$. It decreases if any of these arguments goes to infinity. It has poles at $\xi_i=\eta_j$ and $\xi_i+c=\eta_j$. Finally, the result \eqref{res-Gg1} is correct for $n=1$. Hence, it remains to check that the residues of $\tilde G_n(\gamma)$ in its poles  reduce to $\tilde G_{n-1}$. Then induction over $n$ completes the proof.

We will use the following property of $\Izer_n$:
\be{res-Izer}
\Bigl.\Izer_{n}(\bar x|\bar y)\Bigr|_{x_n\to y_n}= g(x_n,y_n)
f(y_n,\bar y_{n})f(\bar x_{n},x_n) \Izer_{n-1}(\bar x_{n}|\bar y_{n})+{\rm reg},
\ee
where ${\rm reg}$ means the regular part when $x_n\to y_n$.

Let $\eta_n\to \xi_n$. Then the pole in \eqref{def-Gg1} occurs if and only if $\xi_n\in\bar\xi_{\so}$ and
$\eta_n\in\bar\eta_{\so}$. Introducing $\bar\xi_{\so'}=\bar\xi_{\so}\setminus \xi_n$ and $\bar\eta_{\so'}=\bar\eta_{\so}\setminus \eta_n$
we obtain
\begin{multline}\label{Gg-uv-1}
\Bigl.\tilde G_n(\gamma)\Bigr|_{\eta_n\to \xi_n}=\sum (\gamma+n_{\st})
f(\xi_n,\bar\xi_{\st})f(\bar\xi_{\so'},\bar\xi_{\st})f(\bar\eta_{\st},\bar\eta_{\so'})f(\bar\eta_{\st},\eta_n)\\
\times g(\eta_n,\xi_n)f(\bar\eta_{\so'},\eta_n)f(\xi_n,\bar\xi_{\so'})\Izer_{n_{\so}-1}(\bar\eta_{\so'}|\bar\xi_{\so'})
\Izer_{n_{\st}}(\bar\xi_{\st}+c|\bar\eta_{\st})\\
=g(\eta_n,\xi_n)f(\bar\eta_{n},\eta_n)f(\xi_n,\bar\xi_{n})\sum (\gamma+n_{\st})
f(\bar\xi_{\so'},\bar\xi_{\st})f(\bar\eta_{\st},\bar\eta_{\so'})\Izer_{n_{\so}-1}(\bar\eta_{\so'}|\bar\xi_{\so'})
\Izer_{n_{\st}}(\bar\xi_{\st}+c|\bar\eta_{\st})\\
=g(\eta_n,\xi_n)f(\bar\eta_{n},\eta_n)f(\xi_n,\bar\xi_{n})\; \tilde G_{n-1}(\gamma).
\end{multline}

Let now $\xi_n+c=\eta_n$. Then the pole in \eqref{def-Gg1} occurs if and only if $\xi_n\in\bar\xi_{\st}$ and
$\eta_n\in\bar\eta_{\st}$. Introducing $\bar\xi_{\st'}=\bar\xi_{\st}\setminus \xi_n$ and $\bar\eta_{\st'}=\bar\eta_{\st}\setminus \eta_n$ we obtain
\begin{multline}\label{Gg-uv-2}
\Bigl.\tilde G_n(\gamma)\Bigr|_{\eta_n\to \xi_n+c}=\sum (\gamma+n_{\st})
f(\bar\xi_{\so},\xi_{n})f(\bar\xi_{\so},\bar\xi_{\st'})f(\bar\eta_{\st'},\bar\eta_{\so})f(\eta_{n},\bar\eta_{\so})\\
\times h^{-1}(\xi_n,\eta_n)f(\bar\xi_{\st'},\xi_n)f(\eta_n,\bar\eta_{\st'})\Izer_{n_{\so}}(\bar\eta_{\so}|\bar\xi_{\so})
\Izer_{n_{\st}-1}(\bar\xi_{\st'}+c|\bar\eta_{\st'})\\
=h^{-1}(\xi_n,\eta_n)f(\bar\xi_{n},\xi_n)f(\eta_n,\bar\eta_{n})
\sum (\gamma+n_{\st})f(\bar\xi_{\so},\bar\xi_{\st'})f(\bar\eta_{\st'},\bar\eta_{\so})\Izer_{n_{\so}}(\bar\eta_{\so}|\bar\xi_{\so})
\Izer_{n_{\st}-1}(\bar\xi_{\st'}+c|\bar\eta_{\st'})\\
=h^{-1}(\xi_n,\eta_n)f(\bar\xi_{n},\xi_n)f(\eta_n,\bar\eta_{n})\;\tilde G_{n-1}(\gamma+1).
\end{multline}

It is straightforward to check that the r.h.s. of \eqref{res-Gg1} possesses the same recursions at $\eta_n=\xi_n$ and
$\xi_n+c=\eta_n$, and thus, the lemma is proved. Then the statement of lemma~\ref{New-Lemma} immediately follows from lemma~\ref{New-Lemma1}.


\begin{thebibliography}{99}
%
\bibitem{FadST79} L. D. Faddeev, E. K. Sklyanin and L. A. Takhtajan, {\it Quantum Inverse Problem. I.},
 Theor. Math. Phys. {\bf 40} (1979) 688.
 %
\bibitem{FadT79} L. D. Faddeev and L. A. Takhtajan, {\it The quantum method of the inverse problem and the Heisenberg $XYZ$ model},
Usp. Math. Nauk {\bf 34} (1979) 13;  Russian Math. Surveys {\bf 34} (1979) 11 (Engl. transl.).
%
\bibitem{BogIK93L}V. E. Korepin, N. M. Bogoliubov,
A. G. Izergin, {\it Quantum Inverse Scattering Method and Correlation Functions}, Cambridge: Cambridge Univ.
Press, 1993.
%
\bibitem{FadLH96} L. D. Faddeev,  {\it How algebraic Bethe ansatz
works for integrable model},in: Les Houches Lectures, eds A. Connes
et al, North Holland, (1998) 149, \texttt{arXiv:hep-th/9605187}
%
\bibitem{KulRes83}
P. P. Kulish, N. Yu. Reshetikhin,
{\it Diagonalization of $GL(N)$ invariant transfer matrices and quantum $N$-wave system (Lee model)}, J.~Phys.~A:  {\bf 16} (1983) L591.
%
\bibitem{BelPRS12b} S. Belliard, S. Pakuliak, E. Ragoucy, N. A. Slavnov,
{\it The algebraic Bethe ansatz for scalar products in $SU(3)$-invariant integrable
 models}, J. Stat. Mech. (2012) P10017, \texttt{arXiv:1207.0956}.
 %
\bibitem{Kor82}
 V. E. Korepin, {\it Calculation of Norms of Bethe Wave Functions}, Commun. Math. Phys. {\bf 86}
 (1982) 391.
 %
\bibitem{IzeK84}
 A. G. Izergin and V. E. Korepin, {\it The Quantum Inverse Scattering Method Approach to
Correlation Functions}, Commun. Math. Phys. {\bf 94} (1984) 67.
%
\bibitem{KitMST05} N. Kitanine, J.M. Maillet, N.A. Slavnov and V. Terras,
{\it Master equation for spin-spin correlation functions of the
$XXZ$ chain}, Nucl. Phys. B {\bf 712} (2005) 600, \texttt{arXiv:hep-th/0406190}.
%
\bibitem{KitMaiT99}
N. Kitanine, J. M. Maillet, V. Terras, {\it Form factors of the $XXZ$ Heisenberg spin-$1/2$ finite chain},
Nucl. Phys. B {\bf 554} (1999) 647, \texttt{arXiv:math-ph/9807020}.
%
\bibitem{MaiTer00} J. M. Maillet, V. Terras, {\it On the quantum inverse scattering problem}, Nucl. Phys. B {\bf 575} (2000) 627, \texttt{hep-th/9911030}.
%
\bibitem{TarVar93}
V. Tarasov, A. Varchenko, {\it Jackson inte\-gral re\-pre\-sen\-ta\-ti\-ons of so\-lu\-ti\-ons
of the quan\-tized Knizh\-nik--Za\-mo\-lod\-chi\-kov equation},  Algebra and Analysis, {\bf 6} (1994) 90;
St. Petersburg Math. J. {\bf 6} (1995) 275 (Engl. transl.), \texttt{arXiv:hep-th/9311040}.

\bibitem{BelRag08}
S. Belliard, E. Ragoucy,
{\it The nested Bethe ansatz for ``all'' closed spin chains}, J.~Phys.~A {\bf 41} (2008) 295202, \texttt{arXiv:0804.2822}.
%
\bibitem{BelPRS12c} S. Belliard, S. Pakuliak, E. Ragoucy, N. A. Slavnov,
{\it Bethe vectors of $GL(3)$-invariant integrable models}, J. Stat. Mech. (2013) P02020, \texttt{arXiv:1210.0768}.
    %
%
\bibitem{Ize87} A. G. Izergin, {\it Partition function of the six-vertex model in a finite volume},
Dokl. Akad. Nauk SSSR {\bf 297} (1987) 331;
Sov. Phys. Dokl. {\bf 32} (1987) 878 (Engl. transl.).
%
\bibitem{Res86}  N. Yu. Reshetikhin, {\it Calculation of the norm of Bethe vectors in models with $SU(3)$-symmetry}, Zap. Nauchn. Sem. LOMI {\bf 150} (1986) 196;    J. Math. Sci. {\bf 46} (1989 ) 1694 (Engl. transl.).
%
\bibitem{Sla89}
N. A. Slavnov,
{\it Calculation of scalar products of wave functions and form factors in the framework of the algebraic Bethe ansatz}, Theor. Math. Phys. {\bf 79} (1989) 502.
%
\bibitem{Sla07}
N. A. Slavnov,
{\it The algebraic Bethe ansatz and quantum integrable systems}, Usp. Math. Nauk {\bf 62} (2007) 91;
Russian Math. Surveys {\bf 62} (2007) 727 (Engl. transl).
%
\bibitem{KitMT00} N.~Kitanine, J.~M. Maillet, and V.~Terras, {\it Correlation functions of the $XXZ$ Heisenberg
spin-$1/2$ chain in a~magnetic field},  Nucl. Phys. B {\bf 567} (2000) 554, \texttt{arXiv:math-ph/9907019}.
%
\bibitem{KitMST02} N. Kitanine, J.M. Maillet, N.A. Slavnov and V. Terras, {\it Spin-spin correlation functions of the  $XXZ$-$1/2$
Heisenberg chain in a magnetic field}, Nucl. Phys. B {\bf 641} (2002) 487, \texttt{arXiv:hep-th/0201045}.
%
\bibitem{KitKMST09b} N. Kitanine, K.~K. Kozlowski, J. M. Maillet, N. A. Slavnov and V. Terras,
{\it Algebraic Bethe ansatz approach to the asymptotic behavior of correlation functions},
J. Stat. Mech. (2009) P04003, \texttt{arXiv:0808.0227}.
%
\bibitem{GohKS04} F. G\"ohmann, A. Kl\"umper and A. Seel, {\it Integral representations for correlation functions
of the $XXZ$ chain at finite temperature}, J. Phys. A: Math. Gen. {\bf 37} (2004) 7625, \texttt{arXiv:hep-th/0405089}.
%
%
\bibitem{GohKS05} F. G\"ohmann, A. Kl\"umper and A. Seel, {\it Integral representation of the density matrix of the $XXZ$ chain
at finite temperatures}, J. Phys. A: Math. Gen. {\bf 38} (2005) 1833, \texttt{arXiv:cond-mat/0412062}.
%
\bibitem{SeeBGK07} A. Seel, T. Bhattacharyya, F. G\"ohmann, A. Kl\"umper,
{\it A note on the spin-$1/2$ $XXZ$ chain concerning its relation to the Bose gas}, J. Stat. Mech. (2007) P08030, \texttt{arXiv:0705.3569}.
%
\bibitem{CauHM05} J.~S. Caux, J.~M. Maillet, {\it Computation of dynamical correlation functions of Heisenberg chains in a magnetic field},
Phys. Rev. Lett. {\bf 95} (2005) 077201, \texttt{arXiv:cond-mat/0502365}.
%
\bibitem{PerSCHMWA06}
R.~G. Pereira, J. Sirker J, J.~S. Caux, R. Hagemans, J.~M. Maillet, S.~R. White and I.~Affleck,
{\it Dynamical spin structure factor for the anisotropic spin-$1/2$ Heisenberg chain}, Phys. Rev. Lett. {\bf 96} (2006) 257202,
\texttt{arXiv:cond-mat/0603681}.
%
\bibitem{PerSCHMWA07}
R.~G. Pereira, J. Sirker J, J.~S. Caux, R. Hagemans, J.~M. Maillet, S.~R. White and I.~Affleck,
{\it Dynamical structure factor at small $q$ for the $XXZ$ spin-$1/2$ chain},
J. Stat. Mech. (2007) P08022,   \texttt{arXiv:0706.4327}.
%
\bibitem{CauCS07}
J.~S. Caux, P. Calabrese and N. A. Slavnov, {\it One-particle dynamical  correlations in the one-dimensional Bose gas },
J. Stat. Mech. (2007) P01008,  \texttt{arXiv:cond-mat/0611321}.
\end{thebibliography}
\end{document}